\documentclass[11pt,twoside]{article}

\usepackage{asp2006}
\usepackage{epsf}
\usepackage{psfig}
\usepackage{lscape}

\begin{document}
\title{Selecting AGN through variability in SN datasets}   

\author{K. Boutsia}
\affil{INAF - Osservatorio Astronomico di Roma, Via Frascati 33, I-00040 Monteporzio (RM), Italy} 
\author{B. Leibundgut}
\affil{European Southern Observatory, Karl-Schwarzschild-Strasse 2, Garching D-85748, Germany} 
\author{D. Trevese}
\affil{Universit\`a di Roma ``La Sapienza'', Dipartimento di Fisica, P.le A. Moro 2, I-00185 Roma, Italy}
\author{F. Vagnetti}
\affil{Universit\`a di Roma ``Tor Vergata'', Dipartimento di Fisica, via della Ricerca Scientifica 1, I-00133, Roma, Italy}

\begin{abstract} 
Variability is a main property of active galactic nuclei (AGN) and 
it was adopted as a selection criterion using multi epoch surveys conducted 
for the detection of supernovae (SNe). We have used two SN datasets. 
First we selected the AXAF field of the STRESS project, centered in the 
Chandra Deep Field South where, besides the deep X-ray surveys also various 
optical catalogs exist. Our method yielded 132 variable AGN candidates. 
We then extended our method including the dataset of the ESSENCE project 
that has been active for 6 years, producing high quality light curves in 
the R and I bands. We obtained a sample of $\sim$4800 variable sources, 
down to R=22, in the whole 12 deg$^{2}$ ESSENCE field. Among them, 
a subsample of $\sim$500 high priority AGN candidates was created using as 
secondary criterion the shape of the structure function. In a pilot 
spectroscopic run we have confirmed the AGN nature for nearly all 
of our candidates. 
\end{abstract}


\section{Introduction}   

Since the discovery of AGN, variability was established as a main property 
of the population and it was among the first ones to be explored \citep{KBSmith63}. 
The luminosities of AGN have been observed to vary in the whole electromagnetic 
range and the majority of the objects are exhibiting continuum variations of 
about 20\% on timescales of months to years \citep{KBHook94}. From a physical 
point of view, variations can set limits on the size of the central emitting 
region and the differences in the variability properties in the X-ray, optical 
and radio bands provide important information on the underlying structure. The 
mechanism of variability itself is still unknown and a variety of models have 
been proposed \citep{KBTerl92,KBHawk93,KBKawa98}. Thus, the study of the AGN 
variability is very important and can put constraints on the models describing 
the AGN energy source and the AGN structure. 

On the other hand, supernovae (SN) are very powerful cosmological probes and 
their systematic discovery outside the local Universe has led to major scientific 
results, like the confirmation that the Universe is accelerating \citep{KBRiess98,KBPerl99}. 
Such studies require well sampled light curves and large statistical samples which can 
be achieved by monitoring wide areas of the sky to very faint limiting magnitudes. 
This kind of surveys produce huge amounts of data that can be suitable 
also for other scientific studies. For example, given that the time sampling is adequate, 
data gathered during SN searches can be used to detect AGN through variability. 
One of the main purposes of this work is to explore this possibility and create 
suitable tools for the efficient selection of AGN in such databases. The two 
projects that have provided us with their data are: the Southern inTermediate 
Redshift ESO Supernova Search \citep[STRESS,][]{KBBott08} and the 
ESSENCE (Equation of State: SupErNovae trace Cosmic Expansion) survey \citep{KBmikn07}.

\section{AGN in the STRESS survey}

The STRESS survey \citep{KBBott08} includes 16 fields with 
multi-band information. Each of those covers an area of 0.3deg$^2$ 
and has been monitored for 2 years with the ESO/MPI 2.2m telescope.
For this study we choose to use the so-called AXAF field, which is centered at 
$\alpha$=03:32:23.7, $\delta$=-27:55:52 (J2000) and overlaps with various surveys, 
which provide us with further data, such as the COMBO-17 survey \citep{KBWolf03} with 
measurements in 5 wide and 12 narrow filters, resulting in a low resolution 
spectrum, the ESO Imaging Survey \citep[EIS,][]{KBarn01}, the GOODS survey \citep{KBgiav04} 
and the two X-ray surveys, Chandra Deep Field South \citep[CDFS,][]{KBGiac02} and 
the Extended-CDFS survey \citep[ECDFS,][]{KBLeh05}.

For our variability study of the AXAF field we have used 8 epochs 
obtained in the V band, during the period 1999-2001, thus covering 2 years. 
For each source, detected in at least 5 epochs, we have measured the average magnitude 
and its r.m.s. variation which is then compared with a 3$\sigma$ threshold we have 
obtained averaging the r.m.s in bins of magnitude. Details on the calculation 
of the variability threshold can be found in \citet{KBtre08}. This procedure has 
yielded a catalogue of 132 AGN candidates down to V=24~mag. 

Despite all the active surveys in our area, only 31\% of our candidates have 
public spectroscopic information. For this reason we have performed a 
spectroscopic follow up using EMMI at the ESO/NTT (La Silla). We obtained 
low resolution spectra for 27 sources belonging to the bright 
part of our sample (V$<$21.3mag). We now have 55\% of our candidates 
spectroscopically confirmed. The remaining objects are typically fainter than what we have been 
able to observe so far. Based on this dataset a complex picture emerges. Out of 
the 27 sources for which we have obtained spectra, 17 are Broad Line AGNs (BLAGNs), 
1 is a normal galaxy, and a considerable amount (7) are Narrow Emission Line 
Galaxies (NELGs). The remaining two sources are stars. The spectra of all the sources 
we have observed with details about their properties are presented by \citet{KBbout09}.

\begin{figure}[!ht]
\plottwo{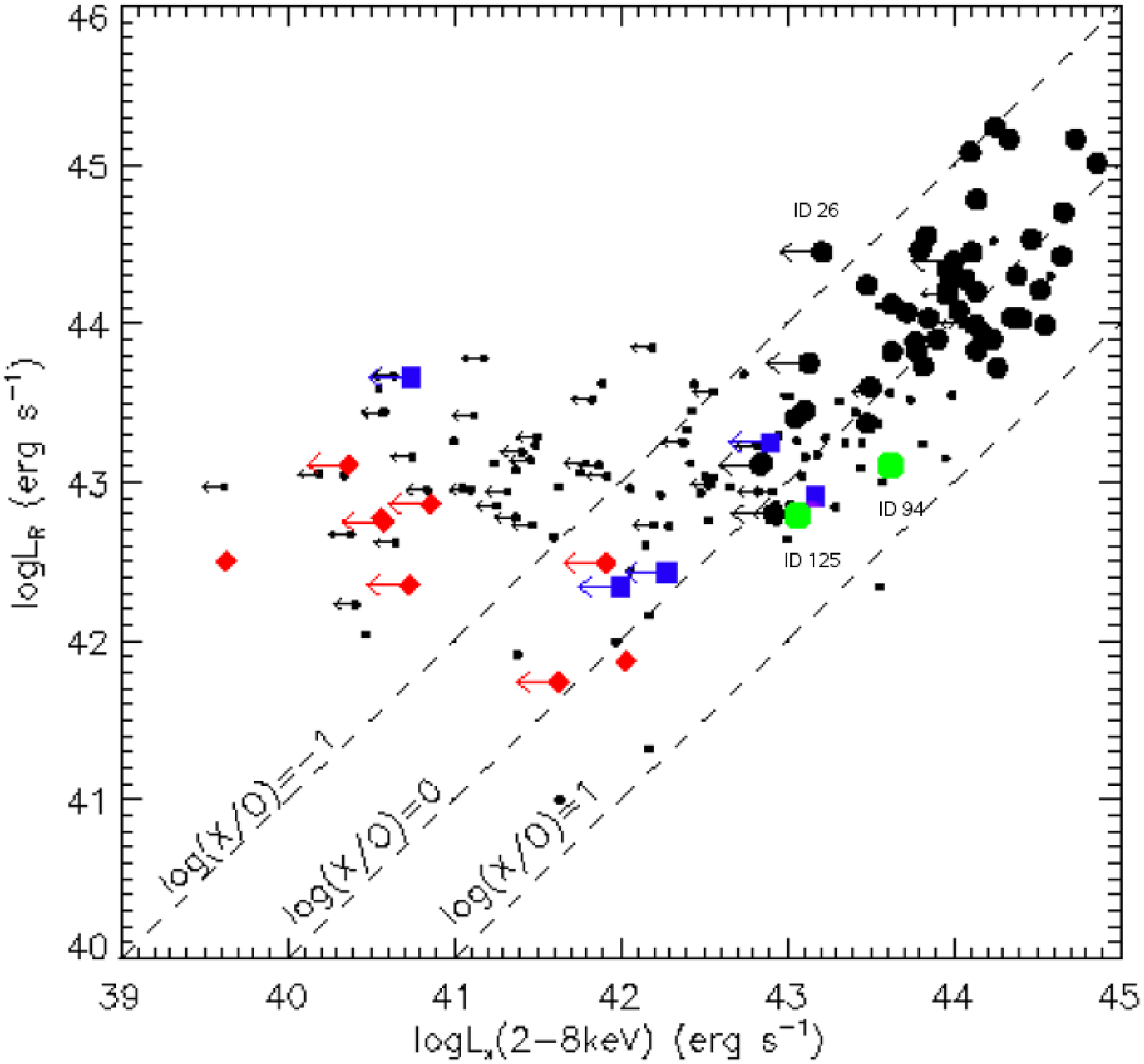}{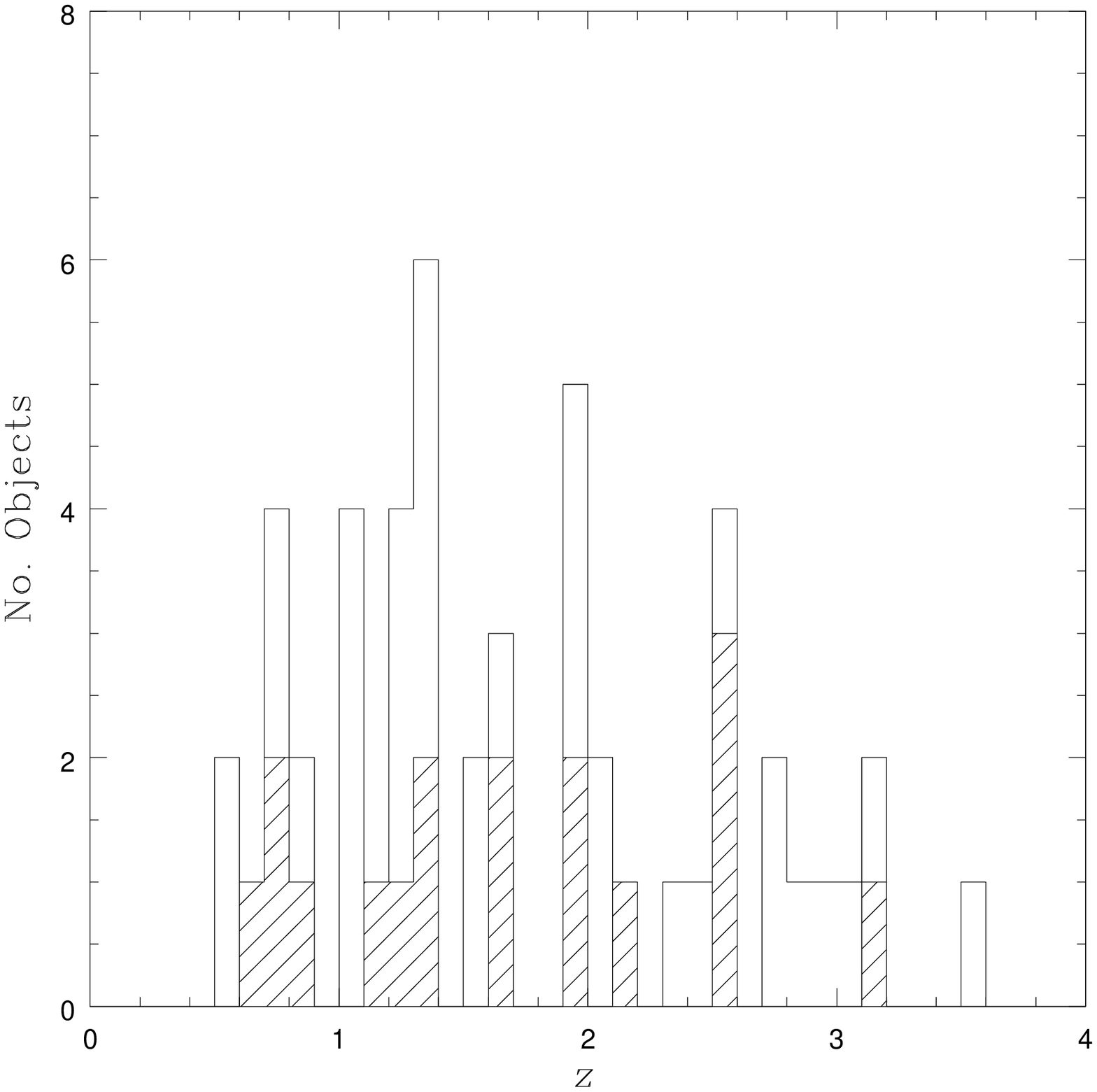}
\caption{Luminosity in the R band vs. luminosity in the hard (2-8keV) X-ray band (left panel). 
Large dots indicate BLAGNs, diamonds represent NELGs and squares are normal galaxies. 
The smaller symbols indicate non variable sources detected in the X-rays in our field. The objects 
with ID 94 and ID 125 indicate sources with X-ray measurements derived from the recent 
2Ms survey \citep{KBluo08}. The source with ID 26 is one of our sources with optical spectra typical 
of BLAGN but undetected by color and X-ray emission. Redshift histogram of the AXAF variable candidates 
(right panel) that have been confirmed as BLAGN based on their optical spectra. The black line refers to all variable 
candidates with known redshift and the shaded section represents the part of the redshifts determined during our campaign.}
\end{figure}

Among the normal galaxies and the NELGs, which make up the low luminosity part 
of our candidates, we may distinguish two groups. A fraction of these sources (7/14) 
displays high variability and their colours ($\ub$ and $\bv$) as well as X-ray to optical ratio 
(X/O), are consistent with AGN (see Fig.1). The other 7 sources, are less variable, have lower X/O 
ratios and their colours are dominated by the host galaxy. According to our analysis, 
these latter sources have properties consistent with Low Luminosity AGN (LLAGN), 
contaminated by the light of the host galaxy. All these sources have extended morphologies 
and would not have been detected by the color technique, that is limited to point like 
sources, nor by their X-ray emission since they are not detected in the hard X-ray 
band (2-8keV) despite the 1Ms exposure time for the CDFS. For the NELGs with the 
necessary lines detected to place them in the diagnostic diagram \citep{KBkew06}, 
we find that they tend to lie in the locus of the composite sources.  

Out of the 65 known BLAGNs in our field, 47 (72\%) are found to display significant 
variability. The confirmed BLAGNs of our sample have an average X/O of 0.55. This 
value is consistent with the X/O of 0.31 obtained by \citet{KBfior03} for optically 
selected samples, while the X-ray selected sources present a ratio of X/O$\sim$1.2. 
This is a further indication that by using variability as a selection technique we probe 
a different part of the AGN population, favouring the identification of X-ray weak 
sources. This fact, in combination with the known correlation between variability 
amplitude and luminosity (in the sense that AGN of lower luminosity show larger variability
amplitudes) makes variability an ideal tool in selecting LLAGNs. Still 45\% of our 
candidates remain without optical spectroscopy because of their faintness. In order 
to better understand the complex LLAGN population, spectroscopy to fainter flux 
limits is needed.

\section{AGN in the ESSENCE survey}

The ESSENCE survey \citep{KBmikn07,KBWV07} has been active for 6 years (2002-2007) and 
was carried out with the Blanco 4m telescope at the Cerro Tololo Inter-American 
Observatory (CTIO). The cadence of the observations was every other night, 
for 20 nights around New Moon, for 3 months per year in the R and I band. 
This resulted in very well sampled light curves for all sources in the 12deg$^2$ field. 

In order to test our variability method we have used only part of the available 
light curves that cover a 2 year period. As it was proven by our previous experience 
in the AXAF field, such timespan is a good compromise between selecting AGN and 
discriminating against supernova outbursts. In such a wide time-baseline with an 
average of 30 epochs per light curve, the variation caused by the SN is well limited 
in time and gets diluted, thus the source does not appear as variable. In fact less 
than 10 of our variable sources resulted to be known SN discovered by the survey. 

The adopted strategy for the ESSENCE dataset is not exactly the same as in the AXAF 
field, mainly because we have used the light curves produced directly by the pipeline 
of the survey for the needs of the project. Here the variability threshold was 
linked to the noise of the data and not the intrinsic variability of the distribution. 
In order to minimize spurious detections, the sources had to show significant variability 
in both bands in order to be classified as candidates. Following this criterion we have 
created a list of $\sim$4800 variable objects down to a magnitude of 22 in the R band. 

Since our light curves are composed by a large number of epochs 
(an average of 30 epochs in each light curve), we may also derive the structure 
function (SF) for each object. The Structure Function (SF) is a method of quantifying 
time variability and according to \cite{KBDeVries05}, it is defined as:

\begin{equation}
\it{S(\tau)} = \{\frac{1}{N(\tau)}\sum_{i<j}[m(t_{i})-m(t_{j})]^{2}\}^{1/2}
\end{equation}

where, N($\tau$) is the number of epochs for which $\it{t_{i}-t_{j}}$=$\tau$ and the 
relative magnitude measurements are summed. We have defined subsamples of candidates, 
depending on the shape of their SF. A flat SF shows that the examined timelag is larger 
than the characteristic time of variability and should indicate sources that are not variable 
or the period of their variability is much shorter than the probed period. In the case 
of AGN, variability is known to increase with time and for the time baseline of the 2 years 
sampled by our light curves, it should result to an ascending SF. Thus, for our spectroscopic 
follow-up we have chosen candidates with ascending or generally non-flat SF. In Fig.2 we 
can see examples of light curves and SFs of sources that were subsequently confirmed as AGN 
by our follow-up. 

\begin{figure}[!ht]
\plottwo{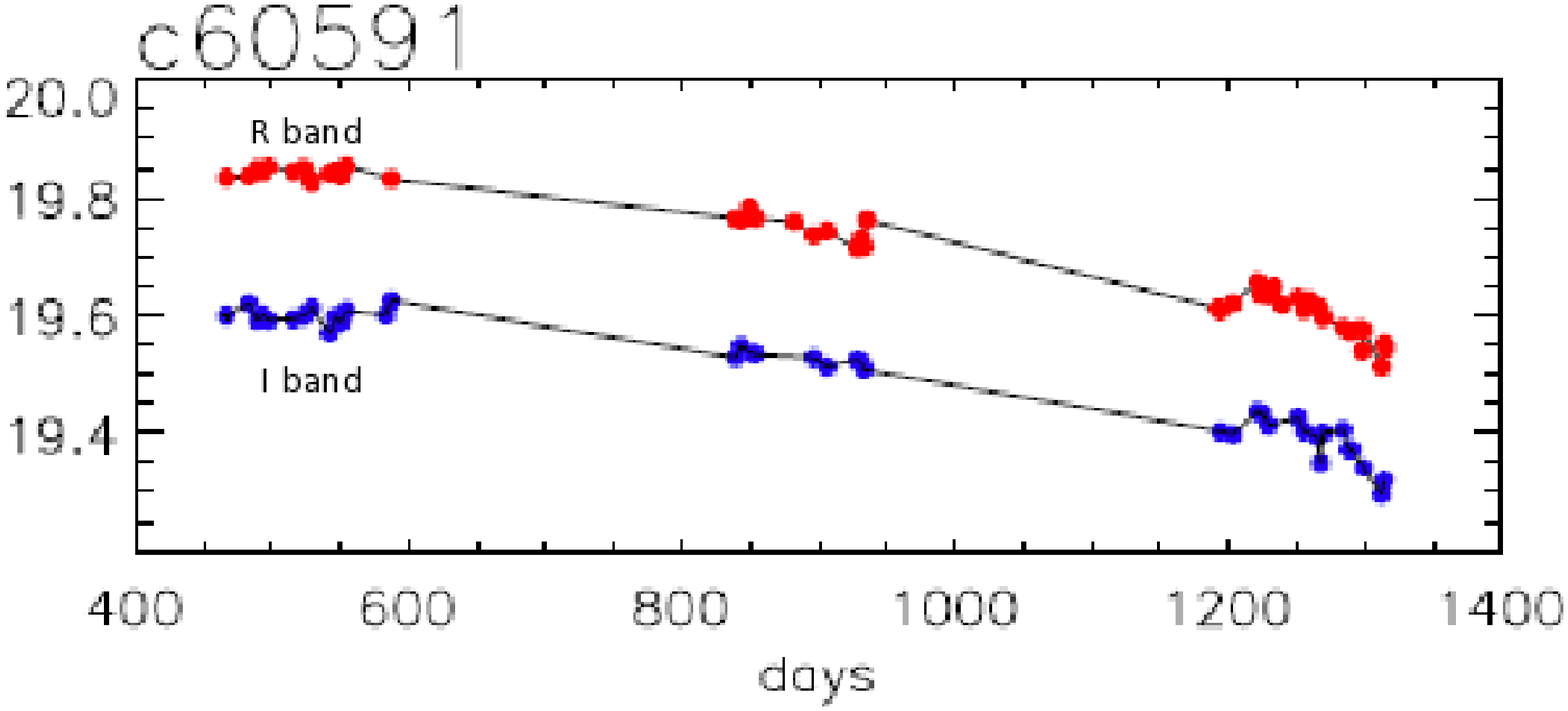}{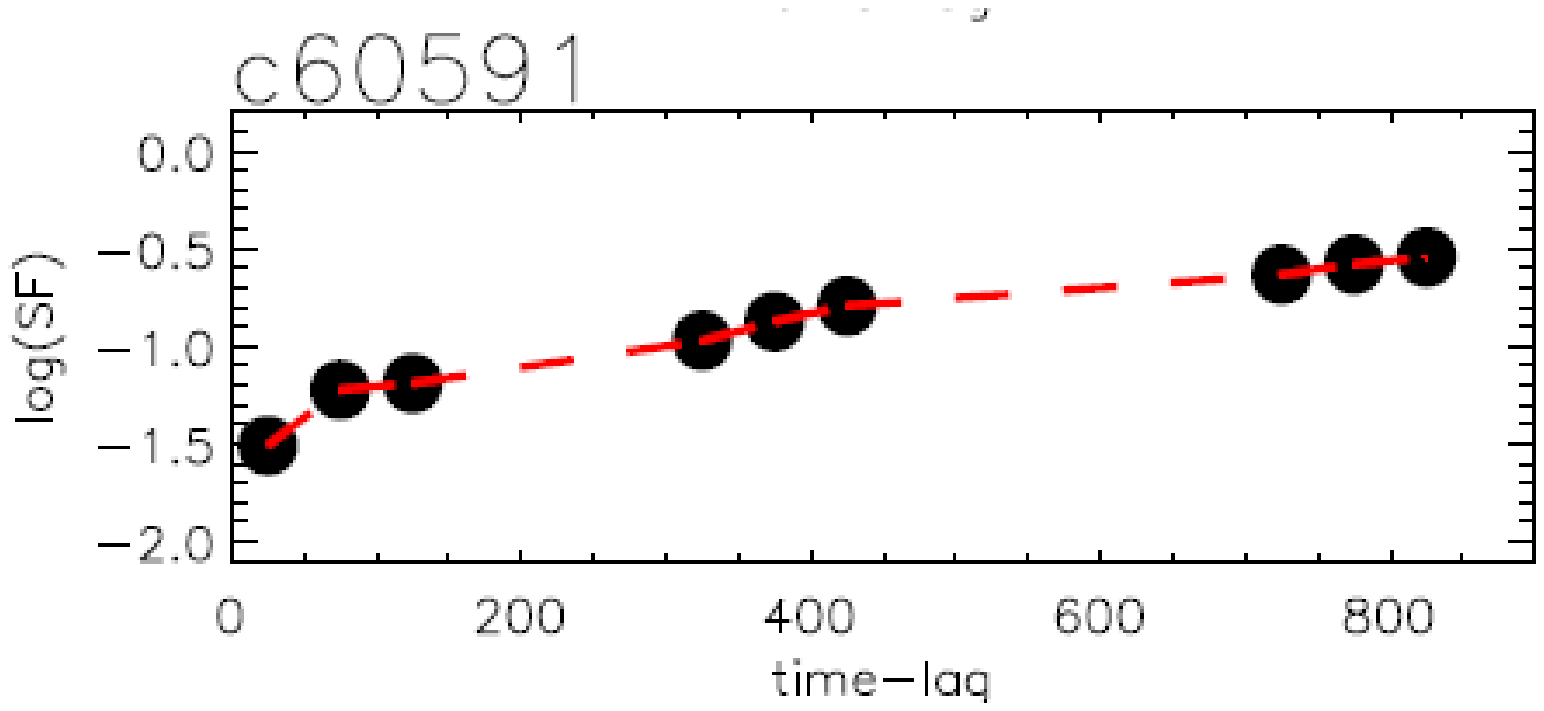}
\plottwo{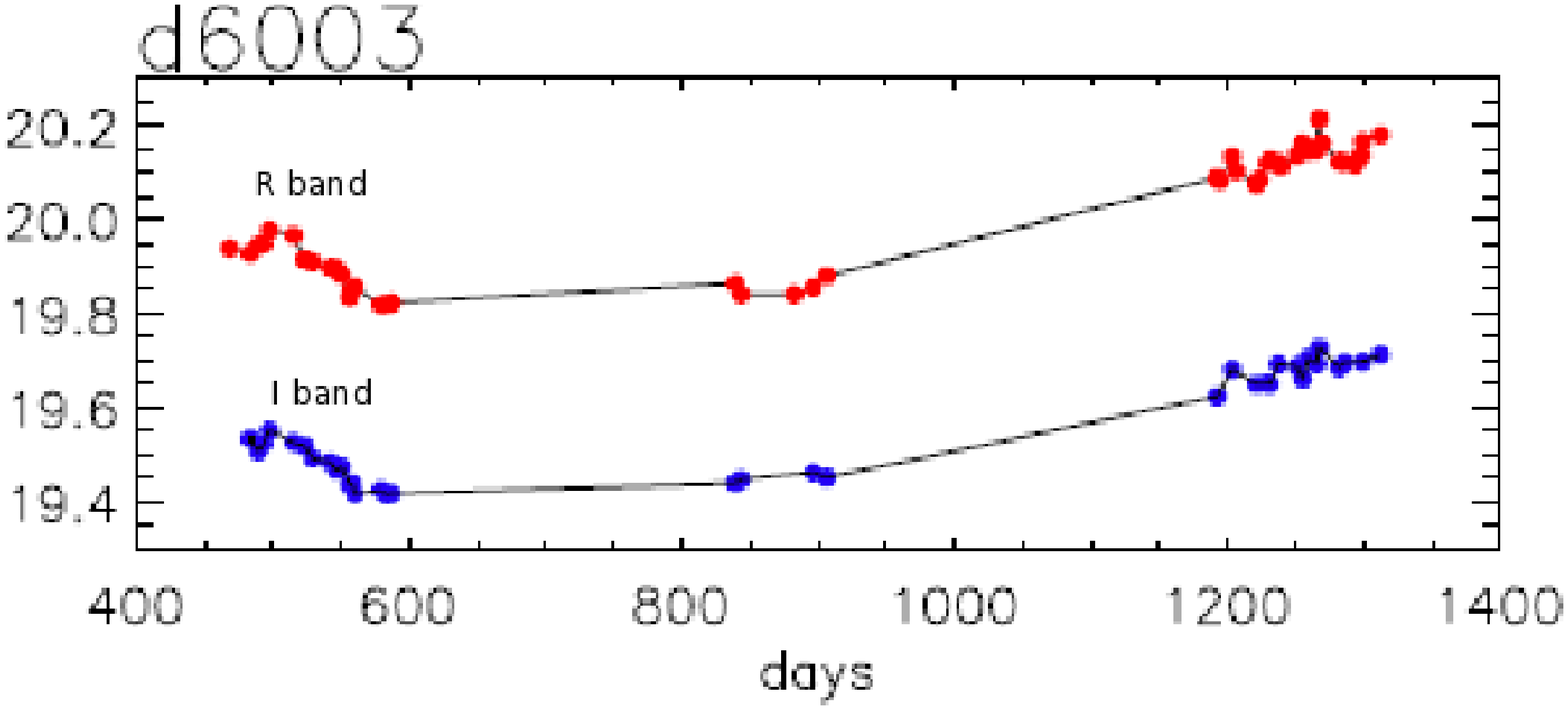}{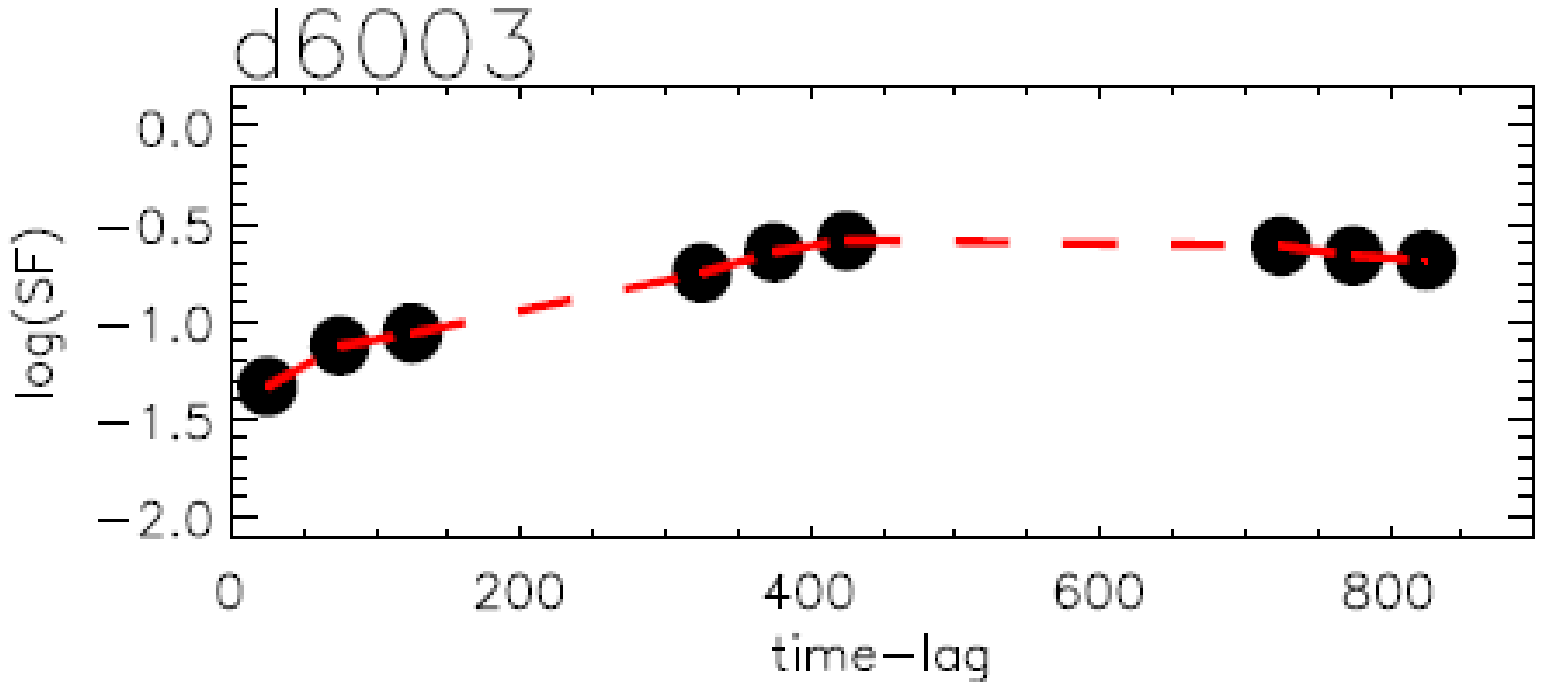}
\caption{Typical light curves in both R and I band (left panel) and binned structure 
functions in the R band (right panel) for two variable candidates that were confirmed 
as BLAGN after our spectroscopic follow-up. Notice the ascending shape of the SF for both 
sources although the shape of their light curves is not comparable.} 
\end{figure}

\begin{figure}[!ht]
\plottwo{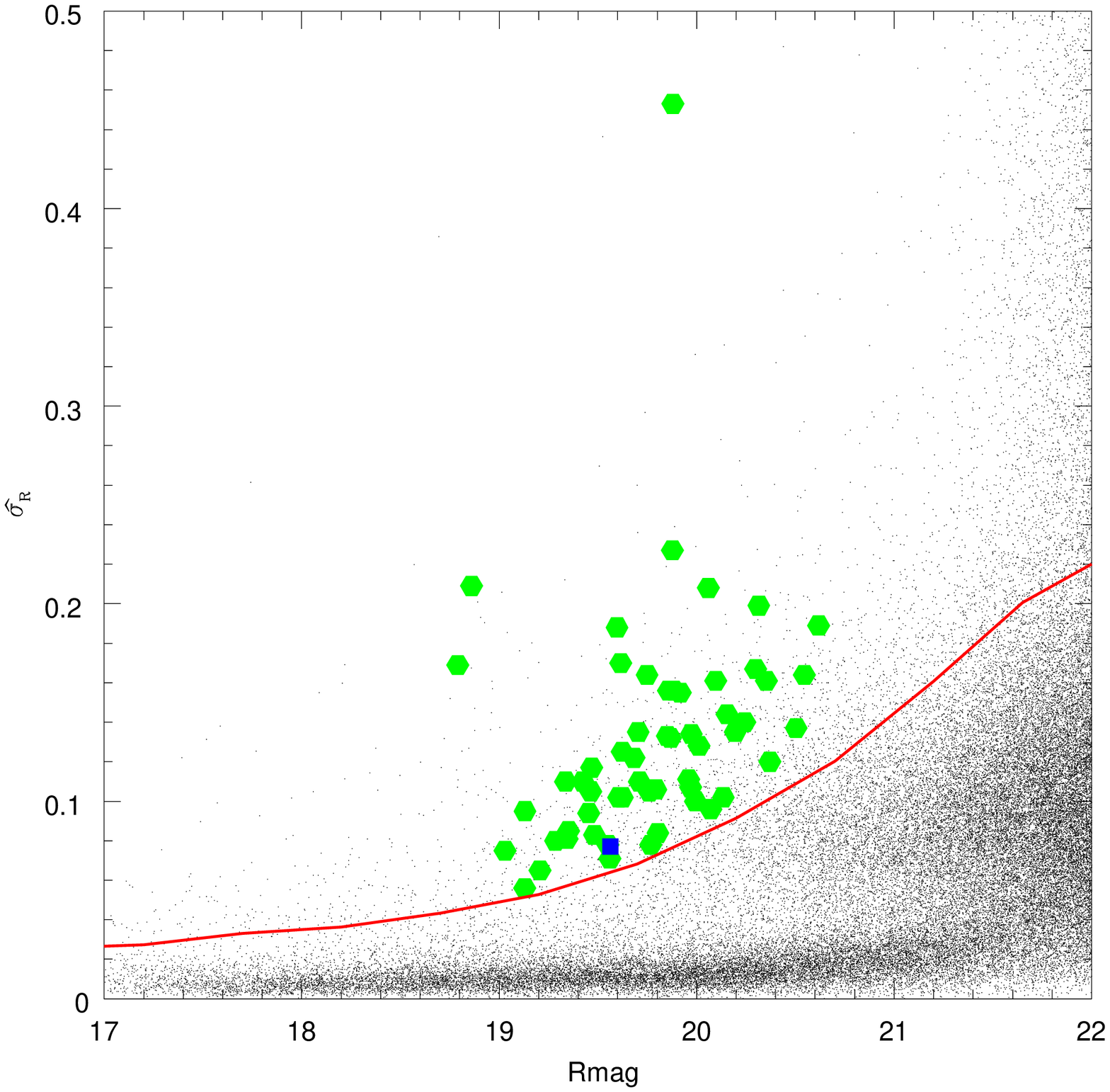}{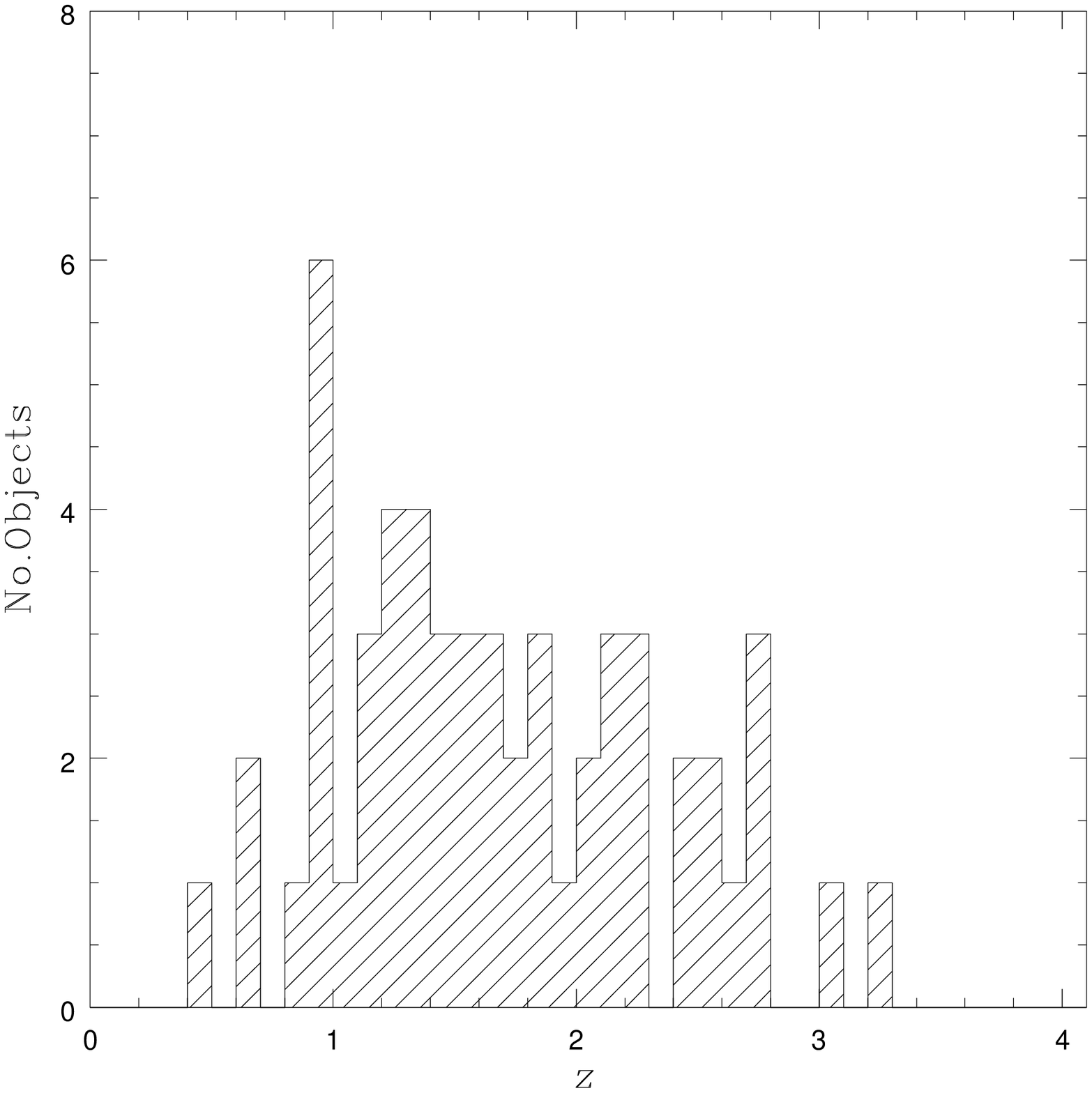}
\caption{Distribution of the variability measurement $\hat{\sigma_{R}}$ versus the R band magnitude 
for all sources that we have observed during our spectroscopic follow-up. The large dots represent 
the confirmed AGN, the square represents the source which is a normal galaxy and the line 
shows the adopted variability threshold. The histogram shows the redshift range of the AGN that were 
observed in the spectroscopic follow-up of our sample.}
\end{figure}

During the same spectroscopic run with EMMI at NTT, we have obtained low resolution 
spectra for 58 sources that belong to the subsample with the ascending SF and their 
magnitudes range  between 18.5 and 20.5 in the R band. 53 (91\%) were confirmed as 
Broad Line AGN and we have a secure redshift determination. 3 sources show only one 
broad emission line and a power law continuum, typical of AGN and although we can not 
claim an accurate redshift for these sources, we can still consider them as $\it{bona}$ 
$\it{fide}$ AGN. This brings our success rate to $\sim$97\%. The remaining sources show 
absorption features of which, one is being recognized as normal galaxy. In Fig.3 we show 
the position of the observed sources in the distribution of their variability versus their 
magnitude. The average redshift of the observed candidates is $<$z$>$=1.40. Details about 
the variability method and the comparison of our sample with other AGN samples existing 
in the same fields will be presented in Boutsia et al. \citetext{in preparation}.

\section{Conclusions}
We have applied a variability selection method to data collected by SN searches 
in order to detect AGN through variability. We have been very successful in 
detecting new AGNs, which had escaped traditional selection techniques and have 
confirmed a large number of already known AGNs in these fields. This proves that 
the AGN field can benefit from such synergic AGN-SN surveys. After a spectroscopic 
follow-up, a considerable fraction of our variable candidates turned out to be ``variable galaxies'' 
with narrow emission lines and properties consistent with LLAGNs diluted by the host galaxy. 
By combining the criterion for variability with a secondary criterion concerning the shape 
of the SF, we created highly reliable AGN samples, since $\sim$97\% of our candidates 
belonging to such a sample was confirmed as BLAGN. In an era that large survey telescopes 
are being developed, such variability studies can give valuable feed-back both for 
determining the strategy of the observations as well as for the development of software 
and pipelines that will allow the scientific community to fully exploit the huge datasets 
that will be produced.



\begin{thebibliography}{}
\bibitem[Arnouts et al.(2001)]{KBarn01}Arnouts, S., Vandame, B., Benoist, C., et al. 2001, A\&A, 379, 740
\bibitem[Botticella et al.(2008)]{KBBott08}Botticella, M.~T., Riello, M., Cappellaro, E., et al. 2008, A\&A, 479, 49
\bibitem[Boutsia et al.(2009)]{KBbout09}Boutsia, K., Leibundgut, B., Trevese, D., \& Vagnetti, F. 2009, A\&A, 497,81
\bibitem[de Vries et al.(2005)]{KBDeVries05}de Vries, W.~H., Becker, R.~H., White, R.~L., \& Loomis, C. 2005, AJ, 129, 615
\bibitem[Giacconi et al.(2002)]{KBGiac02}Giacconi, R., Zirm, A., Wang, J. 2002, ApJS, 139, 369
\bibitem[Giavalisco et al.(2004)]{KBgiav04}Giavalisco, M., Ferguson, H.~C., Koekemoer, A.~M. 2004, ApJ, 600, 93
\bibitem[Fiore et al.(2003)]{KBfior03}Fiore, F., Brusa, M., Cocchia, F. et al. 2003, A\&A, 409, 79
\bibitem[Hawkins(1993)]{KBHawk93}Hawkins, M.~R.~S. 1993, Nature, 366, 242
\bibitem[Hook et al.(1994)]{KBHook94}Hook, I.~M., McMahon, R.~G., Boyle, B.~J., \& Irwin, M.~J. 1994, MNRAS, 268, 305
\bibitem[Kawaguchi et al.(1998)]{KBKawa98}Kawaguchi, T., Mineshige, S., Umemura, M., \& Turner, E.~L. 1998, ApJ, 504, 671
\bibitem[Kewley et al.(2006)]{KBkew06}Kewley, L.~J., Groves, B., Kauffmann, G., \& Heckman, T. 2006, MNRAS, 372, 961
\bibitem[Lehmer et al.(2005)]{KBLeh05}Lehmer, B.D., Brandt, W.N., Alexander, D.M., et al. 2005, ApJS 16, 21
\bibitem[Luo et al.(2008)]{KBluo08}Luo, B., Bauer, F.~E., Brandt, W.~N., et al. 2008, ApJS, 179, 19 
\bibitem[Miknaitis et al.(2007)]{KBmikn07} Miknaitis, G., Pignata, G., Rest, A., et al. 2007, ApJ, 666, 674 
\bibitem[Perl et al.(1999)]{KBPerl99}Perlmutter, S., Turner, M.~S. \& White,M. 1999, Physical Review Letters, 83, 670
\bibitem[Riess et al.(1998)]{KBRiess98}Riess, A.~G., Filippenko, A.~V., Challis, P., et al. 1998, AJ, 116, 1009
\bibitem[Smith et al.(1963)]{KBSmith63}Smith, H.~J.\& Hoffleit, D. 1963, Nature, 198, 650
\bibitem[Terlevich et al.(1992)]{KBTerl92}Terlevich, R., Tenorio-Tagle, G., Franco, J., \& Melnick, J. 1992, MNRAS, 255, 713
\bibitem[Trevese et al.(2008)]{KBtre08}Trevese, D., Boutsia, K., Vagnetti, F. et al. 2008, A\&A, 488, 73
\bibitem[Wolf et al.(2003)]{KBWolf03}Wolf, C., Wisotzki, L., Borch, A., et al. 2003, A\&A, 408, 499
\bibitem[Wood-Vasey et al.(2007)]{KBWV07}Wood-Vasey, W.~M., Miknaitis, G., Stubbs, C.~W., et al. 2007, ApJ, 666, 694
\end{thebibliography}
\end{document}